\documentclass[prd,showpacs,floatfix,nofootinbib,twocolumn,amsmath,amssymb]{revtex4}
\usepackage{graphicx,graphics,color,dcolumn,booktabs,bm}
\usepackage{longtable,lscape}
\usepackage{txfonts}
\usepackage{overpic}
\usepackage{amssymb}
\usepackage{indentfirst}
\usepackage{epsfig}
\usepackage{feynmf}   
\usepackage{epstopdf}   
\usepackage{slashed}  
\usepackage{cases}
\usepackage{color}
\usepackage{multirow}

\usepackage[colorlinks, citecolor=blue,anchorcolor=red,menucolor=red, linkcolor=red,filecolor=red,runcolor=red,urlcolor=blue,frenchlinks=red]{hyperref}
\begin{document}

\title{Production of the charmoniumlike state $Y(4220)$ through the $p\bar{p} \to Y(4220) \pi^0$ reaction} 

\author{Jun-Zhang Wang$^{1,2}$}
\author{Hao Xu$^{1,2}$}
\author{Ju-Jun Xie$^{1,3}$}\email{xiejujun@impcas.ac.cn}
\author{Xiang Liu$^{1,2}$}\email{xiangliu@lzu.edu.cn}
\affiliation{
$^1$Research Center for Hadron
and CSR Physics, Lanzhou University $\&$ Institute of Modern Physics
of CAS,
Lanzhou 730000, China\\
$^2$School of Physical Science and Technology, Lanzhou University,
Lanzhou 730000, China\\
$^3$Institute of Modern Physics, Chinese Academy of Sciences, Lanzhou 730000, China}

\begin{abstract}

The $Y(4220)$ production in the process of $p\bar{p} \to
Y(4220)\pi^0$ is studied within the effective Lagrangian approach.
We first study the $e^+e^-\to Y(4220) \to p\bar{p}\pi^0$ reaction
at center-of-mass energy $4.258$ GeV with the assumption that the
$Y(4220)$ is a $\psi(4S)$ state. We show that the inclusion of
nucleon and nucleon resonance leads to a quite good description of
the experimental data of $\pi^0 p$ and $\pi^0 \bar{p}$ invariant
mass distributions. Then, the total and angular distributions of the
$p\bar{p} \to Y(4220)\pi^0$ reaction are investigated. It is shown
that the nucleon pole is dominant but the contributions from excited
nucleon resonances are non-negligible. The $p\bar{p} \to
Y(4220)\pi^0$ reaction is useful for studying the property of
$Y(4220)$ and our calculated results may be tested at the
forthcoming $\overline{\mbox{P}}$ANDA experiments.
\end{abstract}

 \maketitle

\section{introduction}\label{sec1}

The study of hadron structure and the spectrum of hadron resonances
is one of the most important issues and is attracting much
attention. Recently, more and more charmoniumlike states with
quantum numbers $J^{PC} = 1^{--}$ were discovered, such as
$Y(4008)$, $Y(4230)$, $Y(4260)$, $Y(4360)$, $Y(4630)$, and
$Y(4660)$~\cite{Olive:2016xmw}. In
Refs.~\cite{LlanesEstrada:2005hz,Liu:2007ez,Ding:2007rg,Li:2009zu}
these states were studied under the hypothesis that they are the
conventional vector charmonium states. However, the unusual
properties of some of these states cannot be explained within the
traditional quark
model~\cite{Zhu:2005hp,Chen:2015bft,Chen:2016qju,Liu:2013waa}. In
addition, the discovered vector charmonium states are more than what
the potential quark model predicted. Therefore, further studies for
these charmoniumlike states are needed, such as the hybrid
charmonium, tetraquarks or molecular
states~\cite{Chen:2016qju,Liu:2013waa}.

In 2015, the BESIII and Belle collaborations observed a resonance
structure around $4.23$ GeV with a narrow width in the processes of
$e^+e^- \to \omega \chi_{c0}$~\cite{Ablikim:2014qwy} and $e^+e^- \to
\pi^+ \pi^- \psi(2S)$~\cite{Wang:2014hta}, respectively. Similar
results were also found in $e^+e^-\to \pi^+ \pi^- h_c$
reaction~\cite{Chang-Zheng:2014haa,Ablikim:2013wzq}. The mass and
width of the above observed state (we call it $Y(4230)$ hereafter)
is consistent with the $\psi(4S)$~\cite{Li:2009zu,He:2014xna}.
Later, the decay of $Y(4230) \to \omega\chi_{c0}$ was studied by
treating the $Y(4230)$ as $\psi(4S)$ and the theoretical results are
consistent with the experimental data~\cite{Chen:2014sra}.
Furthermore, a revisit study of the $e^+e^- \to \pi^+\pi^- \psi(2S)$
reaction~\cite{Wang:2014hta} was done in Ref.~\cite{Chen:2015bma} by
including the $\psi(4S)$, $Y(4360)$ and $Y(4660)$ resonances, where
the mass and width for the $\psi(4S)$ are obtained as 4243 MeV and
$16 \pm 31$ MeV, respectively.

In 2017, the BESIII collaboration reported a precise measurement of
$e^+e^-\to \pi^+\pi^-J/\psi$ reaction at center-of-mass (c.m.)
energies between 3.77 to 4.60 GeV~\cite{Ablikim:2016qzw}. Two
resonant structures are observed, one with a mass of
$4222.0\pm3.1\pm1.4$ MeV and a width of $44.1\pm4.3\pm2.0$ MeV and
the other one with a mass of $4320.0\pm10.4\pm7.0$ MeV and a width
of $101.4^{+25.3}_{-19.7}\pm10.2$ MeV. The first resonance,
$Y(4220)$, agrees with the $Y(4230)$ resonance reported by previous
measurements~\cite{Ablikim:2014qwy,Chang-Zheng:2014haa,Ablikim:2013wzq},
while the second resonance is observed for the first time in the
process $e^+ e^- \to \pi^+ \pi^- J/\psi$. The $Y(4220)$ resonance is
also confirmed in the $e^+e^-\to \pi^+\pi^-h_c$
reaction~\cite{BESIII:2016adj}, where the mass $M =
4218.4^{+5.5}_{-4.5} \pm 0.9$ MeV and width $\Gamma =
66.0^{+12.3}_{-8.3} \pm 0.4$ MeV were obtained. On the theoretical
side, one analysis of the recent experimental data on the $e^+e^-\to
\pi^+\pi^-J/\psi$ reaction~\cite{Ablikim:2016qzw} was done in
Ref.~\cite{Chen:2017uof}, where it was found that the $Y(4220)$ must
be introduced to reproduce the detailed data around $\sqrt{s} = 4.2$
GeV of the $e^+  e^- \to \pi^+ \pi^-J/\psi$ and $e^+ e^- \to \pi^+
\pi^- h_c$ cross sections. Furthermore, the possibility of $Y(4220)$
as a charmonium $\psi(4S)$ was also discussed in
Ref.~\cite{Chen:2017uof}. There are also other studies as in
Refs.~\cite{He:2014xna,Dong:1994zj,Li:2009zu}, which provide more
evidence of $Y(4220)$ as $\psi(4S)$.

The study of charmoniumlike states in different production processes
supplies useful information on their properties. As a forthcoming
experiment to study the charmonium and charmoniumlike $XYZ$
states~\cite{Lutz:2009ff}, the Darmstadt ($\overline{\mbox{P}}$ANDA)
experiment at AntiProton and Ion Research (FAIR) is an alternative
platform to study the properties of $Y(4220)$. In this work, we
propose to study the $Y(4220)$ state in the low energy $p\bar{p}$
annihilation within the effective Lagrangian approach. The effective
Lagrangian approach has been widely used to study the charmonium
production in $p\bar{p}$
annihilation~\cite{Gaillard:1982zm,Lundborg:2005am,Barnes:2006ck,Barnes:2007ub,Barnes:2010yb,Lin:2012ru,Liang:2004sd,Zong:2006tg}.
However, only the contribution from exchanged nucleon $N$ was
considered in these previous works. The influence of excited nucleon
resonance $N^*$ may be also
important~\cite{Wiele:2013vla,Xu:2015qqa,Dai:2011yr}. Indeed, in
Ref.~\cite{Xu:2015qqa}, the production of $\psi(3770)$ and
$\psi(3686)$ in $p\bar{p}$ annihilation reaction was studied using
the effective Lagrangian approach, where the contributions from
well-established excited nucleon resonances are also included. In
this work, we extend the work of Ref.~\cite{Xu:2015qqa} to study
$Y(4220)$ production. We study the $e^+e^-\to
p\bar{p}\pi^0$~\cite{Ablikim:2017gtb} to obtain the model
parameters, then we calculate the total and differential cross
sections of $p\bar{p} \to Y(4220) \pi^0$, which may be tested by the
future experiments. Our calculations can provide valuable
information of $Y(4220)$ production at the future
$\overline{\mbox{P}}$ANDA experiments.

This paper is organized as follow. First, the formalism of the
$p\bar{p}\rightarrow Y(4220)\pi^0$ is presented in Sec. \ref{sec2}.
Then in Sec. \ref{sec3}, we fit the process $e^+e^-\rightarrow
Y(4220) \rightarrow p\bar{p}\pi^0$ to experimental data for
obtaining unknown parameters. In Sec. \ref{sec4}, we show the
numerical results and make a detailed discussion. Finally, the paper
ends with the summary in Sec. \ref{sec5}.

\section{Production of $\mathbf{Y(4220)}$ in the $\mathbf{p\bar{p} \rightarrow \pi^0 Y(4220)}$ reaction }\label{sec2}

\begin{figure}[htbp]
\begin{center}
\scalebox{0.50}{\includegraphics{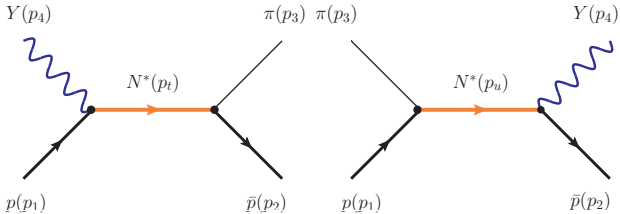}} \caption{The Feynman
diagrams describing the process $p\bar{p}\rightarrow Y(4220)\pi^0$.
Left: $t$-channel exchange; right: $u$-channel exchange. The $Y$
stands for the $Y(4220)$ state and $N^*$ represents nucleon pole and
the excited nuclear resonances. \label{fd1} }
 \end{center}
\end{figure}

The production of charmonium in the low energy $p\bar{p}$
annihilation process can be achieved by exchange intermediated
nucleon states with a light meson emission in the tree level. In
this work, we adopt the effective Lagrangian approach to study the
production of $Y(4220)$ in the $p\bar{p} \rightarrow Y(4220)\pi^0$
reaction, where the contributions from nucleon pole $N$ and five
excited nucleon resonances $N(1440)$, $N(1520)$, $N(1535)$,
$N(1650)$ and $N(1720)$ are considered. The Feynman diagrams of the
$p\bar{p} \rightarrow Y(4220)\pi^0$ are depicted in Fig.~\ref{fd1},
and the interaction vertices of $N^{*}N\pi$ can be described by the
following effective
Lagrangians~\cite{Tsushima:1996xc,Tsushima:1998jz,Zou:2002yy,Ouyang:2009kv,Wu:2009md,Cao:2010km,Cao:2010ji},
\begin{eqnarray}
\mathcal {L}_{\pi NN}&=&-\frac{g_{\pi NN}}{2m_N}\bar{N}\gamma_5\gamma_{\mu}\tau\cdot\partial^{\mu}\pi N, \label{lagup1}\\
\mathcal {L}_{\pi NP_{11}}&=&-\frac{g_{\pi NP_{11}}}{2m_N}\bar{N}\gamma_5\gamma_\mu\tau\cdot\partial^\mu\pi R_{P_{11}}+h.c. ,  \\
\mathcal {L}_{\pi NS_{11}}&=&-g_{\pi NS_{11}}\bar{N}\tau\cdot\pi R_{S_{11}}+h.c.,  \\
\mathcal {L}_{\pi NP_{13}}&=&-\frac{g_{\pi NP_{13}}}{m_N}\bar{N}\tau\cdot\partial_\mu\pi R^\mu_{P_{13}}+h.c.,  \\
\mathcal {L}_{\pi ND_{13}}&=&-\frac{g_{\pi
ND_{13}}}{m_N^2}\bar{N}\gamma_5\gamma^\mu\tau\cdot\partial_\mu\partial_\nu\pi
R^\nu_{D_{13}}+h.c., \label{lagdown1}
\end{eqnarray}
where $g_{N^{*}N\pi}$ is coupling constant of $N^{*}$, $N$ and $\pi$
particle field and $R$ represents the field of excited nucleon and
$\tau$ is Pauli matrix. In addition, the notations $P_{11}$,
$S_{11}$, $P_{13}$ and $D_{13}$ stand for different nucleon states
with $J^{P}=1/2^+, 1/2^-, 3/2^+$ and $ 3/2^- $, respectively.

For the $Y(4220)N^{*}N$ interaction, we adopt the general coupling form
of vector meson and nucleon states ~\cite{Xu:2015qqa}
\begin{eqnarray}
\mathcal {L}_{Y NN}&=&-g_{Y NN}\bar{N}\gamma_{\mu}V^{\mu}N, \label{lagup2}\\
\mathcal {L}_{Y NP_{11}}&=&-g_{Y NP_{11}}\bar{N}\gamma_\mu V^\mu R_{P_{11}}+h.c. ,  \\
\mathcal {L}_{Y NS_{11}}&=&-g_{Y NS_{11}}\bar{N}\gamma_5 \gamma_\mu V^\mu R_{S_{11}}+h.c.,  \\
\mathcal {L}_{Y NP_{13}}&=&-ig_{Y NP_{13}}\bar{N}\gamma_5 V_\mu R^\mu_{P_{13}}+h.c.,  \\
\mathcal {L}_{Y ND_{13}}&=&-g_{Y ND_{13}}\bar{N}V_\mu
R^\mu_{D_{13}}+h.c. , \label{lagdown2}
\end{eqnarray}
where $V^\mu$ stands for the field of the $Y(4220)$ state, and other
symbols are similar with those in the Lagrangian of $N^{*}N\pi$.

According to the above effective Lagrangian densities, the
scattering amplitude of process $p\bar{p}\rightarrow Y(4220)\pi^0$
with different intermediate nucleon states can be derived by
\begin{align}
\mathcal {M}_{\frac{1}{2}^+}=&\frac{g_{\pi NP_{11}} g_{YNP_{11}}}{2m_{N}}\overline{v}(p_2)\epsilon_{\nu}(p_4)
 \bigg[\gamma_5(i\slashed{p}_3)\frac{\slashed{p}_t+m_{N^*}}{t-m_{N^*}^2}\gamma^{\nu}\mathcal{F}(t)  \nonumber\\
  &+\gamma^{\nu}\frac{\slashed{p}_u+m_{N^*}}{u-m_{N^*}^2}\gamma_5(i\slashed{p}_3)\mathcal{F}(u)\bigg]u(p_1),
\label{amplitudeup}\\
\mathcal {M}_{\frac{1}{2}^-}=& g_{\pi N S_{11}}g_{Y N S_{11}}\overline{v}(p_2)\epsilon_{\nu}(p_4)
\nonumber\\
&\times \bigg[\frac{\slashed{p}_t+m_{N^*}}{t-m_{N^*}^2}\gamma_5\gamma^{\nu}\mathcal{F}(t)+\gamma_5\gamma^{\nu}
\frac{\slashed{p}_u+m_{N^*}}{u-m_{N^*}^2}\mathcal{F}(u)\bigg]u(p_1),
\\
\mathcal {M}_{\frac{3}{2}^+}=&i\frac{g_{\pi N P_{13}}g_{Y N P_{13}}}{m_N}\overline{v}(p_2)\epsilon_{\nu}(p_4)
\bigg[(ip_{3\mu})\frac{\slashed{p}_t+m_{N^*}}{t-m_{N^*}^2}G^{\mu\nu}(p_t)   \nonumber\\
&\times\gamma_5 \mathcal{F}(t)+\gamma_5\frac{\slashed{p}_u+m_{N^*}}{u-m_{N^*}^2}G^{\nu\mu}(p_u)
(ip_{3\mu})\mathcal{F}(u)\bigg]u(p_1),
\\
\mathcal {M}_{\frac{3}{2}^-}=&i\frac{g_{\pi N D_{13}}g_{Y N D_{13}}}{m_N^2}\overline{v}(p_2)\epsilon_{\nu}(p_4)
\bigg[\gamma_5(-\slashed{p}_3)(p_{3\mu})\frac{\slashed{p}_t+m_{N^*}}{t-m_{N^*}^2}  \nonumber\\
&\times G^{\mu\nu}(p_t)\mathcal{F}(t)+\frac{\slashed{p}_u+m_{N^*}}{u-m_{N^*}^2}G^{\nu\mu}(p_u)\gamma_5(-\slashed{p}_3)
\nonumber\\
&\times(p_{3\mu})\mathcal{F}(u)\bigg]u(p_1), \label{amplitudedown}
\end{align}
where $t=p_t^2=(p_1-p_4)^2$ and $u=p_u^2=(p_1-p_3)^2$.
$\mathcal{F}(t)$ and $\mathcal{F}(u)$ are form factors for the
$t$-channel and $u$-channel processes, respectively, and
$G^{\mu\nu}(p)$ has the following expression
\begin{eqnarray}
G_{\mu\nu}(p)= -g_{\mu\nu}+\frac{1}{3}\gamma_{\mu}\gamma_{\nu}+\frac{1}{3m_{N^*}}(\gamma_{\mu}p_{\nu}-\gamma_{\nu}p_{\mu})
+\frac{2p_{\mu}p_{\nu}}{3m_{N^*}^2}.
\end{eqnarray}

It is worth noting that the form factors $\mathcal{F}(t/u)$ for the
intermediate off-shell $N$ or $N^*$ state are chosen as that used in
Refs.~\cite{Feuster:1997pq,Haberzettl:1998eq,Yoshimoto:1999dr,Oh:2000zi}
\begin{align}
\mathcal{F}(t/u)=\frac{\Lambda_{N^*}^4}{\Lambda_{N^*}^4+(t/u-m_{N^*}^2)^2},
\end{align}
where the cutoff parameter $\Lambda_{N^*}$ can be parametrized as
\begin{align}
\Lambda_{N^*} = m_{N^*}+\beta \Lambda_{QCD}
\end{align}
with $m_{N^*}$ the mass of intermediate nucleon state and
$\Lambda_{QCD}$=220 MeV, and the free parameter $\beta$ is unknown,
which can be fixed by fitting experimental data.

Finally, based on the above scattering amplitude listed in
Eqs.~(\ref{amplitudeup})-(\ref{amplitudedown}), the general
differential cross section of $p\bar{p}\rightarrow Y(4220)\pi^0$
reads
\begin{eqnarray}
&& \frac{d\sigma}{du} = \frac{1}{16 \pi s
(s-4m^2_p)}\left|\overline{\mathcal{M}}_{p\bar{p} \to Y(4220)\pi^0}\right|^2,
\label{diffential} \\
&& \mathcal{M}_{p\bar{p}\to Y(4220)\pi^0} =
\sum\limits_{N^*}\mathcal{M}_{N^*},
\end{eqnarray}
where $s=(p_1+p_2)^2$ is the invariant mass square of the $p
\bar{p}$ system.

Finally, we determine the coupling constants and the cutoff
parameter $\beta$, by fitting them to the recent BESIII's
measurements of $e^+e^- \to p\bar{p}\pi^0$ reaction at
$\sqrt{s}=4.008-4.6$ GeV~\cite{Ablikim:2017gtb}. In addition, for
the parameter $\beta$, which is directly related to the form factor
of the exchanged nucleon state, it should not be the same for both 
$p\bar{p} \to Y(4220)\pi^0$ and $e^+e^- \to Y(4220) \to
p\bar{p}\pi^0$. Nevertheless, the hadronic form factors in the effective
Lagrangian approach are generally very
phenomenological~\cite{Feuster:1997pq,Haberzettl:1998eq,Yoshimoto:1999dr,Oh:2000zi}
and related available experimental information, so we are
temporarily unable to distinguish the differences between form
factors. We will employ consistent $\beta$ for both processes of
$e^+e^- \to Y(4220) \to p\bar{p}\pi^0$ and $p\bar{p} \to
Y(4220)\pi^0$.

\section{The analysis of  the $\mathbf{e^+e^- \to Y(4220) \to p\bar{p}\pi^0}$ process}\label{sec3}

\begin{figure}[htbp]
\begin{center}
\includegraphics[scale=0.53]{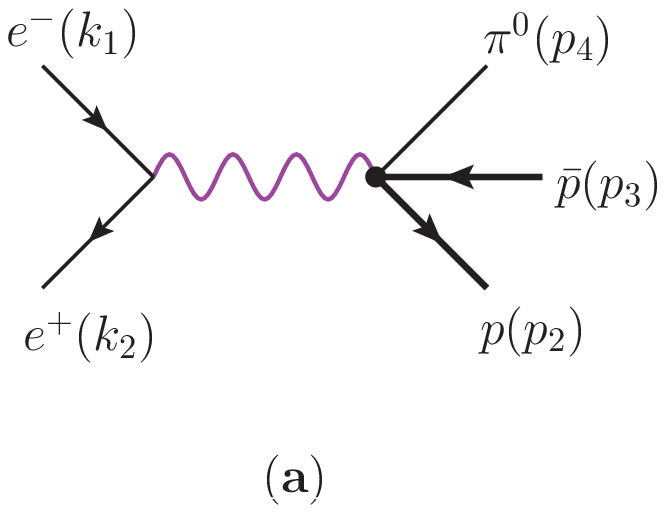}
\includegraphics[scale=0.53]{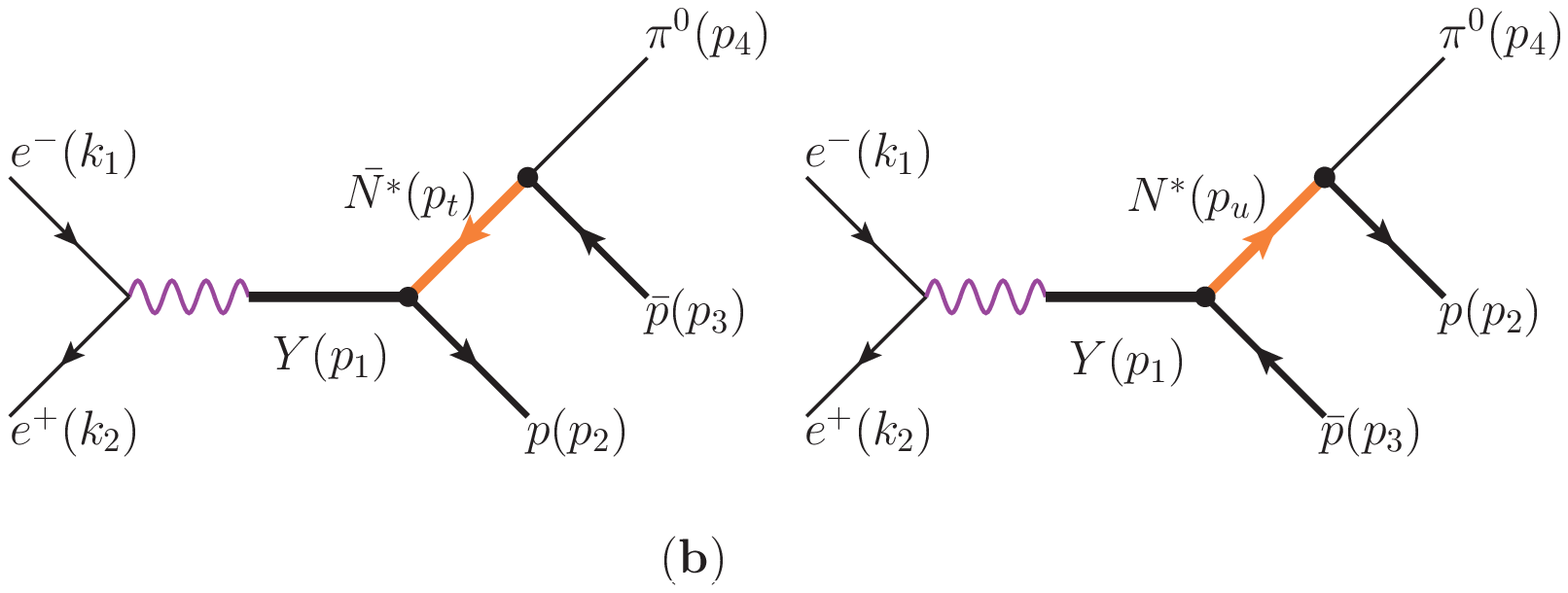}
\caption{ The Feynman diagrams for the process of $e^+e^- \to
p\bar{p}\pi^0$ in the vicinity of $Y(4220)$. Diagram (a)
indicates the contribution of direct process and diagrams (b)
stand for the process $e^+e^- \to Y(4220) \to p\bar{p}\pi^0$.
\label{fd23}}
\end{center}
\end{figure}

The Feynman diagrams of $e^+e^- \to Y(4220) \to p\bar{p}\pi^0$ are
shown in Fig. \ref{fd23} (b). For the $\gamma Y(4220)$ coupling
vertex, we take the vector meson dominant model, where the
corresponding Lagrangian density can be expressed
as~\cite{Lin:2013mka}
\begin{align}
\mathcal{L}_{\gamma V}=\frac{-eM_V^2}{f_V}V_{\mu}A^{\mu},
\end{align}
where $M_V$ and $f_V$ are the mass and decay constant of vector
meson, respectively, and $V_{\mu}$ and $A^{\mu}$ correspond to the
vector meson field and photon field, respectively. The decay
constant $f_V$ can be determined by partial width of $V \to e^+e^-$,
\begin{align}
\frac{e}{f_V}=\left[\frac{3\Gamma_{V\rightarrow e^+e^-}M_V^2}{\alpha
(M^2_V - 4m^2_e)^{3/2}}\right]^{1/2},
\end{align}
where $\alpha={1}/{137}$ is the fine structure constant. Since the
information about the $Y(4220) \to e^+e^-$ is scarce, it is
necessary to rely on theoretical predictions. The value of
$\Gamma_{\psi(4S) \to e^+e^-} = 0.97$ keV is obtained from the color
screened potential model~\cite{Li:2009zu}. With this value we obtain
$e/f_{\psi(4S)}=e/f_{Y(4220)}$=0.0097.

Besides, we consider also the direct background contribution which
is shown in Fig.~\ref{fd23} (a). The direct scattering amplitude of
$e^+e^- \to p\bar{p}\pi^0$ can be written as~\cite{Chen:2010nv}
\begin{align}
\mathcal{M}_{NoR}=g_{NoR}\overline{v}(k_2)e\gamma_{\mu}u(k_1)\frac{1}{s}\overline{u}({p_2})\gamma^{\mu}\gamma_5
v(p_3)\mathcal{F}_{NoR}(s)
\end{align}
with $g_{NoR}$ the coupling constant and,
\begin{eqnarray}
\mathcal{F}_{NoR}(s)=\exp\left[{-a\left(\sqrt{s}-\sum\limits_f m_f\right)^2}\right],
\end{eqnarray}
where $\sum\limits_f m_f$ donates the sum of the mass of final
states of $e^+e^- \to p\bar{p}\pi^0$. Here, the direct term provides
two more free parameters $g_{NoR}$ and $a$.

Then, the total scattering amplitude of $e^+e^- \to p\bar{p}\pi^0$
can be written as
\begin{align}
\mathcal{M}_{e^+e^-\rightarrow p\bar{p}\pi^0}&=\mathcal{M}_{NoR}+\overline{v}(k_2)e\gamma_{\mu}u(k_1)
\frac{-g^{\mu\nu}}{s}eM_{Y}^2/f_{Y(4220)} \nonumber\\
&\quad\times
\frac{-g_{\nu\rho}+p_{Y\nu}p_{Y\rho}/p_Y^2}{s-M_Y^2+iM_Y\Gamma_Y}\mathcal{M}_{Y(4220)\rightarrow
p\bar{p}\pi^0}^{\rho}\label{totamplitude}
\end{align}
with $\mathcal{M}_{Y(4220) \to p\bar{p}\pi^0}^{\rho} =
\sum\limits_{N^*}\mathcal{M}_{N^*}^{\rho}$, where
$\mathcal{M}_{N^*}^{\rho}$ donates the amplitude of each subprocess
of $e^+e^- \to p\bar{p}\pi^0$ with exchanged nuclear state $N^*$ and
these scattering amplitudes can be directly related to Eqs.
(\ref{amplitudeup})-(\ref{amplitudedown}) by making a series of
following substitutions\footnote{See more details in
Ref.~\cite{Xu:2015qqa}.}: $p_1 \rightarrow p_2, p_2\rightarrow p_3,
p_3\rightarrow p_4, p_4\rightarrow p_1$ and $ p_t\rightarrow p_u$, $
p_u\rightarrow p_t$. In addition, the mass $M_{Y(4220)}$ and width
$\Gamma_{Y(4220)}$ are taken as 4222 MeV and 44.1 MeV by the recent
BESIII experiment \cite{Ablikim:2016qzw}, respectively.

The differential cross section of $e^+e^- \to p\bar{p}\pi^0$ reads
\cite{Xie:2015zga}
\begin{align}
d\sigma_{e^+e^-\rightarrow p\bar{p}\pi^0}=\frac{(2\pi)^4 \sum
|\overline{\mathcal{M}}_{e^+e^- \to
p\bar{p}\pi^0}|^2}{4\sqrt{(k_1\cdot k_2)^2}}d\Phi_3 \label{cross1}
\end{align}
with phase space factor
\begin{align}
d\Phi_3=\frac{1}{8(2\pi)^9\sqrt{s}}|\vec{p}_3^{\ast}||\vec{p}_2|d\Omega_3^{\ast}d\Omega_2dM_{\bar{p}\pi},
\label{cross2}
\end{align}
where the overline above $\mathcal{M}$ means the average over the spin
of initial states $e^+$ and $e^-$ and the sum over the spin of the
final state $p$ and $\bar{p}$, and $\vec{p}_3^{\ast}$ stands for the
three-momentum of the antiproton in the center-of-mass frame of the 
$\bar{p}\pi$ system and $M_{\bar{p}\pi}$ is the invariant mass of 
$\bar{p}\pi$.

Next, we perform nine parameters (six coupling constants
$g_{N^*}=g_{Y(4220) NN^*}g_{\pi NN^*}$ for a nucleon pole and five
nucleon resonances, $\beta$, $g_{NoR}$ and $a$) $\chi^2$ fit to the
experimental data of the invariant mass distributions of $p\pi^0$
and $\bar{p}\pi^0$ below 1.85 GeV at $\sqrt{s}$=4.258
GeV~\cite{Ablikim:2017gtb}. It is worth mentioning that the coupling
constants $g_{Y(4220) NN^*}$ and $g_{\pi NN^*}$ always appear by the
form of both products in the scattering amplitude, hence we cannot
temporarily distinguish their separate values. For the measured
cross section data of invariant mass region beyond 1.85 GeV, {{there
is a large contribution}} from higher excited nucleon states, so we
only focus on the experimental data below 1.85 GeV, where the
contributions of nucleon pole $N$ and five excited nucleon
resonances $N(1440)$, $N(1520)$, $N(1535)$, $N(1650)$ and $N(1720)$
as well as direct background are dominant. In addition, compared to
the above nucleon states, the influences of two higher spin nucleon
resonance $N(1675)$ with $J^{P}=\frac{5}{2}^-$ and $N(1680)$ with
$J^{P}=\frac{5}{2}^+$ are found to be suppressed, so they are not
included in this work. The fitted coupling constants $g_{N^*}$ are
shown in Table~\ref{resonant para}, where the inputs for the nucleon
resonances are also shown.

\begin{table}[htbp]
\renewcommand\arraystretch{1.4}
\caption{The resonant parameters of nucleon pole and relevant
excited nucleon states and their corresponding fitted coupling
constants $g_{N^*}$, which is identical to $g_{Y(4220) NN^*}g_{\pi
NN^*}$. The  $m_{N^*}$ and $\Gamma_{N^*}$ are taken the average
value from PDG~\cite{Olive:2016xmw}. } \label{resonant para}
\begin{center}
{\tabcolsep0.09in
\begin{tabular}{lcccrc}
\toprule[1pt] \midrule[1pt]
 $N^*$ &  $J^{P}$  & $m_{N^*}$  (MeV) & $\Gamma_{N^*}$ (MeV)  & $g_{N^*}\,(\times10^{-3})$  \\
\midrule[1pt]
$N(938)$  &   $\frac{1}{2}^+$        & 938             & 0      &  $66.55\pm13.14$           \\
$N(1440)$ &   $\frac{1}{2}^+$        & 1430            & 350    &  $28.13\pm11.42$           \\
$N(1520)$ &   $\frac{3}{2}^-$        & 1515            & 115    &  $13.72\pm3.90$           \\
$N(1535)$ &   $\frac{1}{2}^-$        & 1535            & 150    &  $2.44\pm1.30$           \\
$N(1650)$ &   $\frac{1}{2}^-$        & 1655            & 140    &  $1.60\pm0.88$           \\
$N(1720)$ &   $\frac{3}{2}^+$        & 1720            & 250    &  $1.72\pm0.72$           \\

\midrule[1pt] \bottomrule[1pt]
\end{tabular}
}
\end{center}
\end{table}

\begin{figure}[htbp]
\begin{center}
\scalebox{0.65}{\includegraphics{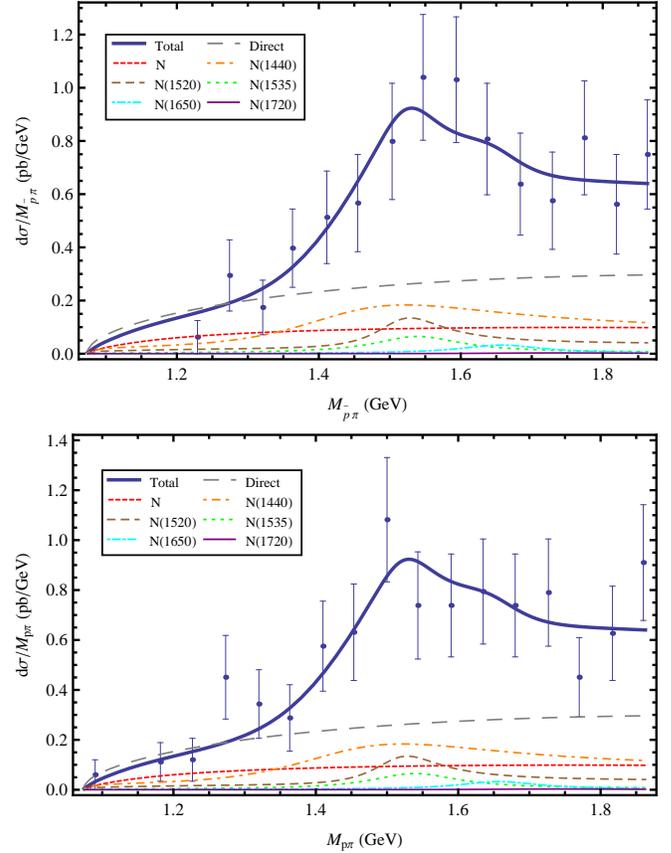}} \caption{The
fitted invariant mass distributions of $\bar{p}\pi^0$ (upper panel)
and $p\pi^0$ (lower panel) of the process $e^+e^- \rightarrow
p\bar{p}\pi^0$ at $\sqrt{s}=4.258$ GeV by utilizing recent BESIII
measurements~\cite{Ablikim:2017gtb}. \label{fittingdata}}
 \end{center}
\end{figure}

The fitted invariant mass distributions of $\bar{p} \pi^0$ and
$p\pi^0$ are shown in Fig. \ref{fittingdata}, where the experimental
data are {{taken from Ref.}} \cite{Ablikim:2017gtb}. The data have been
converted by the results of differential cross section
$d\sigma/M_{\bar{p}\pi^0}$ with measured total cross section
$\sigma_{tot}$=3.08 pb at $\sqrt{s}$=4.258
GeV~\cite{Ablikim:2017gtb}. The fitted $\chi^2/d.o.f = 0.66$ is
obtained with parameters $g_{NoR}=0.87\pm0.34$, $a=0.646\pm0.026$
and $\beta=6.74\pm 3.46$, and the remaining coupling constants
$g_{N^*}$ are all listed in Table~\ref{resonant para}. The
theoretical results can describe the experimental data quite well.
In Fig.~\ref{fittingdata}, the gray-dashed line represents the
contributions of direct background which is shown by Fig.~\ref{fd23}
(a), and the red-dashed line stands for the nucleon pole
contributions, while the other color curves correspond to the
contributions from different excited nucleon resonances. From
Fig.~\ref{fittingdata} we see that the contributions of direct
process are the largest and the contributions from the nucleon pole are
relatively moderate and become stable in the invariant mass region
beyond 1.3 GeV. In addition, an obvious peak around 1.52 GeV is
mainly because of the contributions from $N(1440)$, $N(1520)$ and
$N(1535)$ resonances. Finally, we find that the contributions from
$N(1650)$ and $N(1720)$ resonances are relatively small, especially
for the $N(1720)$ state whose resonant peak structure is nearly
disappeared.

\section{The total cross sections and angular distributions of $\mathbf{p\bar{p}\rightarrow \pi^0 Y(4220)}$}\label{sec4}

In this section, based on the studies of $e^+ e^- \to Y(4220) \to p
\bar{p} \pi^0$, we investigate the production of $Y(4220)$ in the
$p\bar{p} \to Y(4220)\pi^0$ reaction.

\begin{figure}[htbp]
\begin{center}
\scalebox{0.75}{\includegraphics{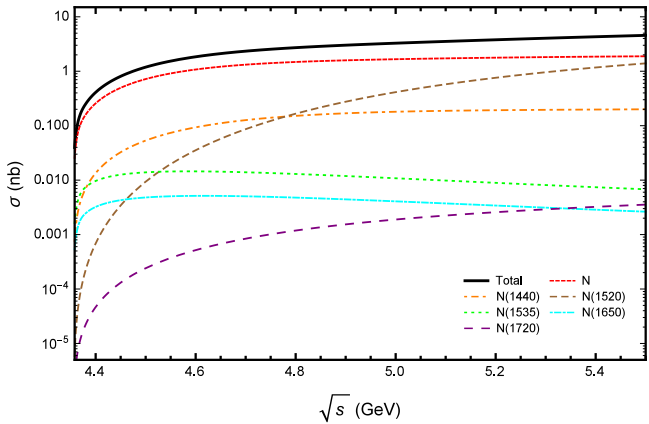}} \caption{
The total cross sections of $p\bar{p} \to Y(4220)\pi^0$ as a
function of $\sqrt{s}$. \label{totalcross}}
 \end{center}
\end{figure}

With those coupling constant $g_{N^*}$ and cutoff $\beta$ have been
determined by fitting process $e^+e^- \to p\bar{p}\pi^0$, we can
calculate the total cross sections and angular distributions of the 
$p\bar{p} \to Y(4220) \pi^0$ reaction. The numerical results of the
total cross sections as a function of the c.m. energy $\sqrt{s}$ are
shown in Fig.~\ref{totalcross}, where the black-solid line stands
for the total results, while the red-dashed curve and other color
line correspond to the contributions from the nucleon pole and different
excited nuclear states, respectively. From Fig.~\ref{totalcross}, it
is found that the contribution from nucleon pole is dominant and the
total cross section is increasing rapidly at small $\sqrt{s}$ and it
tends to flat at high $\sqrt{s}$, where the contributions from
excited nucleon resonances are also important. In addition, it is
quite interesting that the results of the excited nucleon states
with different quantum numbers $J^{P}$ perform very different
behavior. The trend of $N(1440)$ is exactly the same with the nucleon pole
and the curve of $N(1535)$ and $N(1650)$ with $J^{P}=\frac{1}{2}^-$
showing a decreasing trend, while others increase very fast, and the
contribution from $N(1520)$ resonance becomes more significant at
$\sqrt{s} > 5.2$ GeV, which is comparable with the nucleon pole.

\begin{figure}[htbp]
\begin{center}
\scalebox{0.75}{\includegraphics{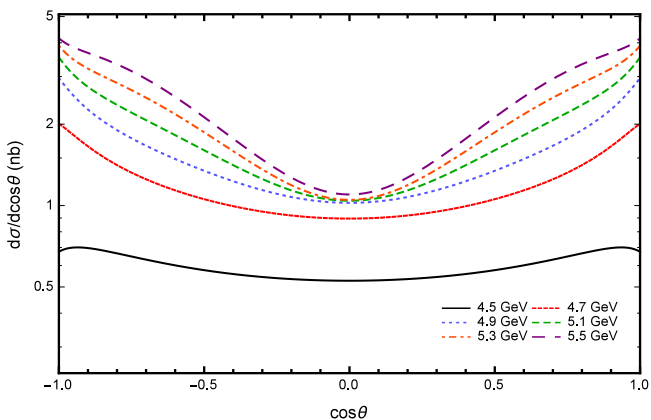}} \caption{ The
angular distributions of $p\bar{p} \to Y(4220)\pi^0$ with the
contributions from both nucleon pole and excited nucleon resonances.
\label{angulardis1}}
 \end{center}
\end{figure}

The angular distributions, $d\sigma/d{\rm cos}\theta =
2|\vec{p}_1||\vec{p}_3|d\sigma/du$, of the $p\bar{p} \to Y(4220)\pi^0$
reaction are also calculated and the results are shown in
Fig.~\ref{angulardis1}. The $\theta$ is the angle of outgoing
$\pi^0$ relative to the ingoing proton beam in the c.m. frame. In
Fig.~\ref{angulardis1}, the different color curves stand for six
c.m. energy points from 4.5 to 5.5 GeV. It is find that the angular
distributions are symmetry at forward and backward angles, which is
because the contributions from $u$-channel and $t$-channel have the same
weights. Meanwhile, the differential cross section of $p\bar{p} \to
Y(4220)\pi^0$ near $\theta=90^{\circ}$ remains a relatively stable
value when $\sqrt{s}$ is larger than $4.9$ GeV. {{On the other hand,
there are small bump structures at ${\rm cos}\theta=0.94$ and
$-0.94$ when the c.m. energy is set as 4.5 GeV}}.

\begin{figure}[htbp]
\begin{center}
\scalebox{0.75}{\includegraphics{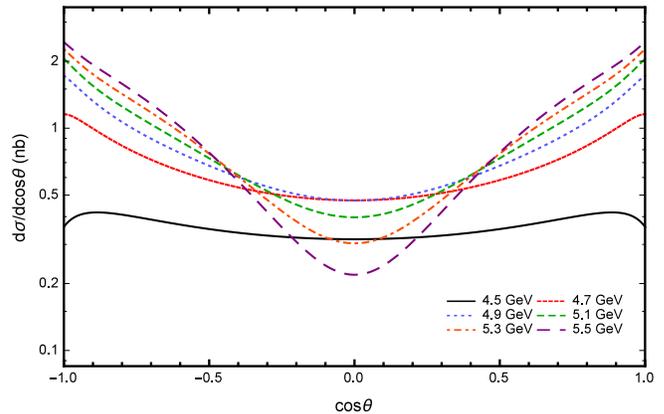}} \caption{ The
angular distributions of $p\bar{p} \to Y(4220)\pi^0$ with the
contributions from the nucleon pole alone. \label{angulardis2}}
 \end{center}
\end{figure}

To show the effect from the nucleon pole, we perform a calculation
of the angular distributions from only the nucleon pole and the
numerical results are given in Fig.~\ref{angulardis2}. From
Figs.~\ref{angulardis1} and \ref{angulardis2}, we can see that the
shapes of the angular distributions of the nucleon pole and the total
contributions are much different. We hope that this feature may be
used to determine the reaction mechanisms of $\bar{p}p \to Y(4220)
\pi^0$ reaction by future experimental analysis.

\section{Summary}\label{sec5}

Recently, the BESIII collaboration made progress on searching for
vector charmoniumlike states. {{By analyzing the $e^+e^-\rightarrow
J/\psi \pi^+\pi^-$}} reaction at center-of-mass energy
$\sqrt{s}=3.77-4.60$  GeV \cite{Ablikim:2016qzw}, BESIII indicated
that there exist two charmoniumlike structures $Y(4220)$ and
$Y(4330)$. $Y(4220)$ also exists in the $e^+e^-\to h_c\pi^+\pi^-$
process \cite{BESIII:2016adj}. Stimulated by the observation of
charmoniumlike state $Y(4220)$, in this work we study the
production of newly observed charmoniumlike state $Y(4220)$
\cite{Ablikim:2016qzw,BESIII:2016adj} via the low energy
$p\bar{p}\to Y(4220)\pi^0$ reaction. For presenting the information
of total cross section and the corresponding angular distributions
of the $p\bar{p}\rightarrow Y(4220)\pi^0$ process which is useful to
experimental study, we adopt the effective Lagrangian approach, where we
consider intermediate nucleon resonance contributions to
$p\bar{p}\rightarrow Y(4220)\pi^0$ and $Y(4220)$ is treated as
charmonium $\psi(4S)$~\cite{Chen:2014sra}. In the concrete
calculation, some input parameters like coupling constants and
cutoff can be fixed by the data of $e^+e^-\rightarrow p\bar{p}\pi^0$
at $\sqrt{s}=4.258$ GeV suggested in Ref.~\cite{Xu:2015qqa}. Our
study also indicates that the nucleon pole has a dominant contribution
in the $p\bar{p}\rightarrow Y(4220)\pi^0$ reaction. Additionally, we
also find that the contributions from other five well-established
excited nucleon resonances $N(1440)$, $N(1520)$, $N(1535)$,
$N(1650)$ and $N(1720)$ cannot be ignored, since these intermediate
excited nuclear states have a sizable contribution to the behavior of the
cross section. Thus, the forthcoming $\overline{\mbox{P}}$ANDA
experiment has potential to carry out the study of production of
charmoniumlike state $Y(4220)$ through the $p\bar{p} \to Y(4220)
\pi^0$ reaction, which may provide new way to establish $Y(4220)$ in the 
future.

Until now, the vector charmonium family has not been well
established \cite{Olive:2016xmw}. It is obvious that the present
work is useful for informing additional experimental efforts to
fully define the spectroscopy of the $J/\psi$ system in the $4-5$
GeV mass region. Here, the $\overline{\mbox{P}}$ANDA experiment will be
an important platform to achieve this goal.

\section*{ACKNOWLEDGMENTS}

This work is partly supported by the National Natural Science
Foundation of China under Grants No. 11175073  and
No. 11475227. It is also supported by the Youth Innovation Promotion
Association CAS (No. 2016367) and the Fundamental Research Funds for the Central Universities. Xiang Liu is also supported in part by the National Program for Support of Top-notch Young Professionals.


\begin{thebibliography}{99}


\bibitem{Olive:2016xmw}
  C.~Patrignani {\it et al.} [Particle Data Group],
  Review of Particle Physics,
  \href{http://dx.doi.org/10.1088/1674-1137/40/10/100001}{Chin.\ Phys.\ C {\bf 40}, 100001 (2016)}.

\bibitem{LlanesEstrada:2005hz}
  F.~J.~Llanes-Estrada,
  Y(4260) and possible charmonium assignment,
  \href{http://dx.doi.org/10.1103/PhysRevD.72.031503}{Phys.\ Rev.\ D {\bf 72}, 031503 (2005)}.

\bibitem{Liu:2007ez}
  X.~Liu,
  Understanding the newly observed Y(4008) by Belle,
  \href{http://dx.doi.org/10.1140/epjc/s10052-008-0551-4}{Eur.\ Phys.\ J.\ C {\bf 54}, 471 (2008)}.

\bibitem{Ding:2007rg}
  G.~J.~Ding, J.~J.~Zhu and M.~L.~Yan,
  Canonical Charmonium Interpretation for Y(4360) and Y(4660),
  \href{http://dx.doi.org/10.1103/PhysRevD.77.014033}{Phys.\ Rev.\ D {\bf 77}, 014033 (2008)}.


\bibitem{Li:2009zu}
  B.~Q.~Li and K.~T.~Chao,
  Higher Charmonia and X,Y,Z states with Screened Potential,
  \href{http://dx.doi.org/10.1103/PhysRevD.79.094004}{Phys.\ Rev.\ D {\bf 79}, 094004 (2009)}.


\bibitem{Zhu:2005hp}
  S.~L.~Zhu,
  The Possible interpretations of Y(4260),
  \href{http://dx.doi.org/10.1016/j.physletb.2005.08.068}{Phys.\ Lett.\ B {\bf 625}, 212 (2005)}.


\bibitem{Chen:2015bft}
  D.~Y.~Chen, X.~Liu, X.~Q.~Li and H.~W.~Ke,
  Unified Fano-like interference picture for charmoniumlike states $Y(4008)$, $Y(4260)$ and $Y(4360)$,
  \href{http://dx.doi.org/10.1103/PhysRevD.93.014011}{Phys.\ Rev.\ D {\bf 93}, 014011 (2016)}.

\bibitem{Chen:2016qju}
  H.~X.~Chen, W.~Chen, X.~Liu and S.~L.~Zhu,
  The hidden-charm pentaquark and tetraquark states,
  \href{http://dx.doi.org/10.1016/j.physrep.2016.05.004}{Phys.\ Rept.\  {\bf 639}, 1 (2016)}.

\bibitem{Liu:2013waa}
  X.~Liu,
  An overview of $XYZ$ new particles,
  \href{http://dx.doi.org/10.1007/s11434-014-0407-2}{Chin.\ Sci.\ Bull.\  {\bf 59}, 3815 (2014)}.

\bibitem{Ablikim:2014qwy}
  M.~Ablikim {\it et al.} [BESIII Collaboration],
  Study of $e^+e^-\to\omega\chi_{cJ}$ at center-of-mass energies from 4.21 to 4.42 GeV,
  \href{http://dx.doi.org/10.1103/PhysRevLett.114.092003}{Phys.\ Rev.\ Lett.\  {\bf 114}, 092003 (2015)}.

\bibitem{Wang:2014hta}
  X.~L.~Wang {\it et al.} [Belle Collaboration],
  Measurement of $e^+e^- \to \pi^+\pi^-\psi(2S)$ via Initial State Radiation at Belle,
  \href{http://dx.doi.org/10.1103/PhysRevD.91.112007}{Phys.\ Rev.\ D {\bf 91}, 112007 (2015)}.


\bibitem{Chang-Zheng:2014haa}
  C.~Z.~Yuan,
  Evidence for resonant structures in $e^{+}e^{-} \to \pi^{+}\pi^{-}h_c$,
  \href{http://dx.doi.org/10.1088/1674-1137/38/4/043001}{Chin.\ Phys.\ C {\bf 38}, 043001 (2014)}.



\bibitem{Ablikim:2013wzq}
  M.~Ablikim {\it et al.} [BESIII Collaboration],
  Observation of a Charged Charmoniumlike Structure $Z_c$(4020) and Search for the $Z_c$(3900) in $e^+e^- \to \pi^+\pi^-h_c$,
  \href{http://dx.doi.org/10.1103/PhysRevLett.111.242001}{Phys.\ Rev.\ Lett.\  {\bf 111}, 242001 (2013)}.



\bibitem{He:2014xna}
  L.~P.~He, D.~Y.~Chen, X.~Liu and T.~Matsuki,
  Prediction of a missing higher charmonium around 4.26 GeV in $J/\psi$ family,
  \href{http://dx.doi.org/10.1140/epjc/s10052-014-3208-5}{Eur.\ Phys.\ J.\ C {\bf 74}, 3208 (2014)}.

\bibitem{Chen:2014sra}
  D.~Y.~Chen, X.~Liu and T.~Matsuki,
  Observation of $e^+e^-\to \chi_{c0}\omega$ and missing higher charmonium $\psi(4S)$,
  \href{http://dx.doi.org/10.1103/PhysRevD.91.094023}{Phys.\ Rev.\ D {\bf 91}, 094023 (2015)}.

\bibitem{Chen:2015bma}
  D.~Y.~Chen, X.~Liu and T.~Matsuki,
  Search for missing $\psi(4S)$ in the $e^+e^-\to \pi^+\pi^-\psi(2S)$ process,
  \href{http://dx.doi.org/10.1103/PhysRevD.93.034028}{Phys.\ Rev.\ D {\bf 93}, 034028 (2016)}.



\bibitem{Ablikim:2016qzw}
  M.~Ablikim {\it et al.} [BESIII Collaboration],
  Precise measurement of the $e^+e^-\to \pi^+\pi^-J/\psi$ cross section at center-of-mass energies from 3.77 to 4.60 GeV,
  \href{http://dx.doi.org/10.1103/PhysRevLett.118.092001}{Phys.\ Rev.\ Lett.\  {\bf 118}, 092001 (2017)}.





\bibitem{BESIII:2016adj}
  M.~Ablikim {\it et al.} [BESIII Collaboration],
  Evidence of Two Resonant Structures in $e^+ e^- \to \pi^+ \pi^- h_c$,
  \href{http://dx.doi.org/10.1103/PhysRevLett.118.092002}{Phys.\ Rev.\ Lett.\  {\bf 118}, 092002 (2017)}.


\bibitem{Chen:2017uof}
  D.~Y.~Chen, X.~Liu and T.~Matsuki,
  Interference effect as resonance killer of newly observed charmoniumlike states $Y(4320)$ and $Y(4390)$,
  \href{https://arxiv.org/pdf/1708.01954.pdf}{arXiv:1708.01954 [hep-ph]}.

\bibitem{Dong:1994zj}
  Y.~B.~Dong, Y.~W.~Yu, Z.~Y.~Zhang and P.~N.~Shen,
  Leptonic decay of charmonium,
  \href{https://journals.aps.org/prd/pdf/10.1103/PhysRevD.49.1642}{Phys.\ Rev.\ D {\bf 49}, 1642 (1994)}.


\bibitem{Lutz:2009ff}
  M.~F.~M.~Lutz {\it et al.} [PANDA Collaboration],
  Physics Performance Report for PANDA: Strong Interaction Studies with Antiprotons,
  \href{https://arxiv.org/pdf/0903.3905.pdf}{arXiv:0903.3905 [hep-ex]}.



\bibitem{Gaillard:1982zm}
  M.~K.~Gaillard, L.~Maiani and R.~Petronzio,
  Soft Pion Emission in $p \bar{p}$ Resonance Formation,
  \href{http://dx.doi.org/10.1016/0370-2693(82)91044-9}{Phys.\ Lett.\  {\bf 110B}, 489 (1982)}.

\bibitem{Lundborg:2005am}
  A.~Lundborg, T.~Barnes and U.~Wiedner,
  Charmonium production in $p \bar{p}$ annihilation: Estimating cross sections from decay widths,
  \href{http://dx.doi.org/10.1103/PhysRevD.73.096003}{Phys.\ Rev.\ D {\bf 73}, 096003 (2006)}.

\bibitem{Barnes:2006ck}
  T.~Barnes and X.~Li,
  Associated Charmonium Production in Low Energy $p \bar{p}$ Annihilation,
  \href{http://dx.doi.org/10.1103/PhysRevD.75.054018}{Phys.\ Rev.\ D {\bf 75}, 054018 (2007)}.

\bibitem{Barnes:2007ub}
  T.~Barnes, X.~Li and W.~Roberts,
  Evidence for a $J/\psi p\bar{p}$ Pauli Strong Coupling?,
  \href{http://dx.doi.org/10.1103/PhysRevD.77.056001}{Phys.\ Rev.\ D {\bf 77}, 056001 (2008)}.

\bibitem{Barnes:2010yb}
  T.~Barnes, X.~Li and W.~Roberts,
  A Meson Emission Model of $\Psi \to N \bar{N}$ m Charmonium Strong Decays,
  \href{http://dx.doi.org/10.1103/PhysRevD.81.034025}{Phys.\ Rev.\ D {\bf 81}, 034025 (2010)}.

\bibitem{Lin:2012ru}
  Q.~Y.~Lin, H.~S.~Xu and X.~Liu,
  Revisiting the production of charmonium plus a light meson at PANDA,
  \href{http://dx.doi.org/10.1103/PhysRevD.86.034007}{Phys.\ Rev.\ D {\bf 86}, 034007 (2012)}.

\bibitem{Liang:2004sd}
  W.~H.~Liang, P.~N.~Shen, B.~S.~Zou and A.~Faessler,
  Nucleon pole contributions in $J / \psi \to N \bar{N}\pi$, $p\bar{p} \eta$, $p\bar{p} \eta^\prime$ and $p \bar{p} \omega$ decays,
  \href{http://dx.doi.org/10.1140/epja/i2004-10007-y}{Eur.\ Phys.\ J.\ A {\bf 21}, 487 (2004)}.

\bibitem{Zong:2006tg}
  Y.~Y.~Zong, P.~N.~Shen, B.~S.~Zou, J.~F.~Liu and W.~H.~Liang,
  Re-study of nucleon pole contribution in $J/\psi \to N \bar{N} \pi$ decay,
  \href{http://dx.doi.org/10.1088/0253-6102/46/3/021}{Commun.\ Theor.\ Phys.\  {\bf 46}, 507 (2006)}.

\bibitem{Wiele:2013vla}
  J.~Van de Wiele and S.~Ong,
  Study of associated charmonium J/$\Psi$ production in $\bar{p}p \to \pi^0$ + J/$\Psi$,
  \href{http://dx.doi.org/10.1140/epjc/s10052-013-2640-2}{Eur.\ Phys.\ J.\ C {\bf 73}, 2640 (2013)}.

\bibitem{Dai:2011yr}
  J.~P.~Dai, P.~N.~Shen, J.~J.~Xie and B.~S.~Zou,
  $J/\psi \rightarrow p\bar{p}\phi$ decay in the isobar resonance model,
  \href{http://dx.doi.org/10.1103/PhysRevD.85.014011}{Phys.\ Rev.\ D {\bf 85}, 014011 (2012)}.

\bibitem{Xu:2015qqa}
  H.~Xu, J.~J.~Xie and X.~Liu,
  Implication of the observed $e^{+}e^{-}\rightarrow p{\bar{p}}\pi ^0$ for studying the $p{\bar{p}}\rightarrow \psi (3770)\pi ^0$ process,
  \href{http://dx.doi.org/10.1140/epjc/s10052-016-4054-4}{Eur.\ Phys.\ J.\ C {\bf 76}, 192 (2016)}.



\bibitem{Ablikim:2017gtb}
  M.~Ablikim {\it et al.} [BESIII Collaboration],
  Cross section measurements of $e^{+}e^{-}\rightarrow p \bar{p} \pi^{0}$ at center-of-mass energies between 4.008 and 4.600 GeV,
  \href{https://arxiv.org/pdf/1701.04198.pdf}{arXiv:1701.04198 [hep-ex]}.


\bibitem{Tsushima:1996xc}
  K.~Tsushima, A.~Sibirtsev and A.~W.~Thomas,
  Resonance model study of strangeness production in $p p$ collisions,
  \href{http://dx.doi.org/10.1016/S0370-2693(96)01391-3}{Phys.\ Lett.\ B {\bf 390}, 29 (1997)}.

\bibitem{Tsushima:1998jz}
  K.~Tsushima, A.~Sibirtsev, A.~W.~Thomas and G.~Q.~Li,
  Resonance model study of kaon production in baryon baryon reactions for heavy ion collisions,
  \href{http://dx.doi.org/10.1103/PhysRevC.61.029903}{Phys.\ Rev.\ C {\bf 59}, 369 (1999)}
  \href{http://dx.doi.org/10.1103/PhysRevC.59.369}{Erratum: [Phys.\ Rev.\ C {\bf 61}, 029903 (2000)]}.

\bibitem{Zou:2002yy}
  B.~S.~Zou and F.~Hussain,
  Covariant L-S scheme for the effective $N^*NM$ couplings,
  \href{http://dx.doi.org/10.1103/PhysRevC.67.015204}{Phys.\ Rev.\ C {\bf 67}, 015204 (2003)}.

\bibitem{Ouyang:2009kv}
  Z.~Ouyang, J.~J.~Xie, B.~S.~Zou and H.~S.~Xu,
  Theoretical study on $pp \to p n \pi^+$ reaction at medium energies,
  \href{http://dx.doi.org/10.1142/S0218301309012306}{Int.\ J.\ Mod.\ Phys.\ E {\bf 18}, 281 (2009)}.

\bibitem{Wu:2009md}
  J.~J.~Wu, Z.~Ouyang and B.~S.~Zou,
  Proposal for Studying $N^*$ Resonances with $\bar{p} p \to \bar{p} n \pi^+$ Reaction,
  \href{http://dx.doi.org/10.1103/PhysRevC.80.045211}{Phys.\ Rev.\ C {\bf 80}, 045211 (2009)}.

\bibitem{Cao:2010km}
  X.~Cao, B.~S.~Zou and H.~S.~Xu,
  Phenomenological analysis of the double pion production in nucleon-nucleon collisions up to 2.2 GeV,
  \href{http://dx.doi.org/10.1103/PhysRevC.81.065201}{Phys.\ Rev.\ C {\bf 81}, 065201 (2010)}.

\bibitem{Cao:2010ji}
  X.~Cao, B.~S.~Zou and H.~S.~Xu,
  Phenomenological study on the $\bar p N\to \bar NN\pi\pi$ reactions,
  \href{http://dx.doi.org/10.1016/j.nuclphysa.2011.05.094}{Nucl.\ Phys.\ A {\bf 861}, 23 (2011)}.









\bibitem{Feuster:1997pq}
  T.~Feuster and U.~Mosel,
  A Unitary model for meson nucleon scattering,
  \href{http://dx.doi.org/10.1103/PhysRevC.58.457}{Phys.\ Rev.\ C {\bf 58}, 457 (1998)}.


\bibitem{Haberzettl:1998eq}
  H.~Haberzettl, C.~Bennhold, T.~Mart and T.~Feuster,
  Gauge-invariant tree-level photoproduction amplitudes with form factors,
  \href{http://dx.doi.org/10.1103/PhysRevC.58.R40}{Phys.\ Rev.\ C {\bf 58}, R40 (1998)}.

\bibitem{Yoshimoto:1999dr}
  T.~Yoshimoto, T.~Sato, M.~Arima and T.~S.~H.~Lee,
  Dynamical test of constituent quark models with pi N reactions,
  \href{http://dx.doi.org/10.1103/PhysRevC.61.065203}{Phys.\ Rev.\ C {\bf 61}, 065203 (2000)}.

\bibitem{Oh:2000zi}
  Y.~s.~Oh, A.~I.~Titov and T.~S.~H.~Lee,
  Nucleon resonances in omega photoproduction,
  \href{http://dx.doi.org/10.1103/PhysRevC.63.025201}{Phys.\ Rev.\ C {\bf 63}, 025201 (2001)}.


\bibitem{Lin:2013mka}
  Q.~Y.~Lin, X.~Liu and H.~S.~Xu,
  Charged charmoniumlike state $Z_c(3900)^¡À$ via meson photoproduction,
  \href{http://dx.doi.org/10.1103/PhysRevD.88.114009}{Phys.\ Rev.\ D {\bf 88}, 114009 (2013)}.


\bibitem{Chen:2010nv}
  D.~Y.~Chen, J.~He and X.~Liu,
  Nonresonant explanation for the Y(4260) structure observed in the $e^+e^-\to J/\psi\pi^+\pi^-$ process,
  \href{http://dx.doi.org/10.1103/PhysRevD.83.054021}{Phys.\ Rev.\ D {\bf 83}, 054021 (2011)}.

\bibitem{Xie:2015zga}
  J.~J.~Xie, Y.~B.~Dong and X.~Cao,
  Role of the $\Lambda^+_c(2940)$ in the $\pi^- p \to D^- D^0 p$ reaction close to threshold,
  \href{http://dx.doi.org/10.1103/PhysRevD.92.034029}{Phys.\ Rev.\ D {\bf 92}, 034029 (2015)}.






\end{thebibliography}
\end{document}